\documentclass[pre,amssymb,floatfix]{revtex4}
\usepackage{graphicx,bm} \usepackage{subfigure}
\newcommand{\be}{\begin{equation}} \newcommand{\ee}{\end{equation}}
\newcommand{\bea}{\begin{eqnarray}} \newcommand{\eea}{\end{eqnarray}}
\newcommand{\angintphi}{\int_{-\frac{\pi}{2}}^{\frac{\pi}{2}}d\phi\frac{\cos\phi}2}
\newcommand{\angintpsi}{\int_{-\frac{\pi}{2}}^{\frac{\pi}{2}}d\psi\frac{\cos\psi}2}
\newcommand{\n}{\tilde{n}} 
\begin{document}

\title{Systematic Density Expansion of the Lyapunov Exponents for a
Two-dimensional Random Lorentz Gas}

\author{H.\ V.\ Kruis, Debabrata Panja, Henk van Beijeren}
\affiliation{Institute for Theoretical Physics, University of Utrecht,
Postbus 80.195, 3508 TD Utrecht, The Netherlands}

\begin{abstract}

\noindent We study the Lyapunov exponents of a two-dimensional, random
Lorentz gas at low density. The positive Lyapunov exponent may be
obtained either by a direct analysis of the dynamics, or by the use of
kinetic theory methods. To leading orders in the density of scatterers
it is of the form $A_{0}\tilde{n}\ln\tilde{n}+B_{0}\tilde{n}$, where
$A_{0}$ and $B_{0}$ are known constants and $\tilde{n}$ is the number
density of scatterers expressed in dimensionless units. In this paper,
we find that through order $(\tilde{n}^{2})$, the positive Lyapunov
exponent is of the form
$A_{0}\tilde{n}\ln\tilde{n}+B_{0}\tilde{n}+A_{1}\tilde{n}^{2}\ln\tilde{n}
+B_{1}\tilde{n}^{2}$. Explicit numerical values of the new constants
$A_{1}$ and $B_{1}$ are obtained by means of a systematic
analysis. This takes into account, up to $O(\tilde{n}^{2})$, the
effects of {\it all\/} possible trajectories in two versions of the
model; in one version overlapping scatterer configurations are allowed
and in the other they are not.

\vspace{2mm}
\noindent {\bf Keywords:} Lyapunov exponent, Lorentz gas, density
expansion.

\end{abstract}

\maketitle

\section{Introduction}

The Lorentz gas is a model consisting of a set of scatterers that
are fixed in space, together with a moving point particle (or a
cloud of mutually non-interacting point-particles) undergoing
collisions with the scatterers. Here we will consider two variants
of the two-dimensional version, where the scatterers are fixed
hard disks, placed at random in the plane and the collisions of
the point particle with the scatterers are elastic and specular.
In the first version the positions of the scatterers are
completely random, so different scatterers may overlap each other
(this corresponds to the case of point scatterers with a moving
particle of circular shape).  In the second version the scatterers
may not overlap each other, but each configuration satisfying this
constraint has equal a priori weight (this corresponds to a
hard-sphere interaction between the scatterers). The Lorentz Gas
has proved to be very useful for studying the general relations
existing between dynamical systems theory and the non-equilibrium
properties of many body systems. Explicit examples of such
relationships encompass the escape rate formalism of Gaspard and
Nicolis \cite {GasNic,D_cup_book} as well as the Gaussian
thermostat formalism developed by Evans and co-workers
\cite{gaussEv,D_cup_book}  and by Hoover and co-workers \cite
{gaussHoov,D_cup_book}. The Lorentz gas is known to be chaotic due
to the convex shape of the boundaries of the scatterers. Its
chaotic properties can be analyzed in fair detail, at least if the
system is sufficiently dilute, in other words, the average
distance between neighboring scatterers is large compared to their
radius. For the model where the scatterers are placed on a
periodic lattice (the Sinai billiard) Sinai has shown mixing and
ergodicity \cite{Sinai_rms_70} and demonstrated that on large time
and length scales the motion of the point particle is diffusive
\cite{commentSinai}. Already much earlier Krylov
\cite{Krylov_book_pup} conjectured that to leading order the
positive Lyapunov exponent is of the form $\tilde{n}\ln\tilde{n}$,
with $\tilde{n}\equiv na^2$ proportional to the density of
scatterers. Subsequently Van Beijeren {\it et al.}
\cite{vBD_prl_95,vBD_prl_96,vBLD_pre_98} carried out kinetic
theory calculations yielding explicit expressions for the Lyapunov
exponents of a disordered Lorentz gas (random configuration of
scatterers) to leading orders, i.e.\ {\it up to} $O(\tilde{n})$ in
the scatterer density. These studies confirmed Krylov's conjecture
and provided explicit values for the constants $A_{0}$ and $B_{0}$
appearing in the leading terms
$A_{0}\tilde{n}\ln\tilde{n}+B_{0}\tilde{n}$ of the expansion of
the largest Lyapunov exponent in terms of the scatterer density.
The kinetic theory methods used for obtaining this result, employ
an averaging over all allowed configurations of the scatterers.

Up to {\em quadratic} order in the density of scatterers, the positive
Lyapunov exponent can be calculated by similar kinetic theory methods,
refining the averaging procedure described above, such that it (i)
takes into account the effects of non-overlap of the scatterers (for
the non-overlapping scatterer model) and (ii) accounts for the most
important effects of correlated collisions. The additional
contributions to the positive Lyapunov exponent resulting from (i) and
(ii) yield the positive Lyapunov exponent correct through order
$\tilde{n}^{2}$. The purpose of this paper is to present a {\it
systematic\,} analysis to calculate these additional contributions and
to give explicit values for the new coefficients appearing in the
expression
$A_{0}\tilde{n}\ln\tilde{n}+B_{0}\tilde{n}+A_{1}\tilde{n}^{2}\ln\tilde{n}+B_{1}\tilde{n}^{2}$
for the positive Lyapunov exponent up to this order. This is done in
two steps: in the first step, we calculate the positive Lyapunov
exponent up to $O(\tilde{n}^2)$, assuming that the collisions between
the point particle with the scatterers are uncorrelated. This embodies
that the particle does not encounter any scatterer more than once,
plus that all previous knowledge on the presence and absence of
scatterers on the track of the point particle is ignored.  However, in
the non-overlapping scatterer model the influence of the non-overlap
condition on the collision frequency is taken into account (for the
overlapping model no corrections are needed for this at the given
orders in $\n$). In the second step, we consider (ii), and calculate
corrections to the Lyapunov exponents due to correlated collision
sequences in which the point particle either encounters the same
scatterer more than once, or its scattering probability is enhanced or
suppressed by the knowledge resulting from previous collisions, or the
absence thereof.  This includes the immediate suppression of
collisions as result of the non-overlap condition.

The structure of this paper is as follows: in Sec.\ II, we briefly
discuss the general theory of the Lyapunov exponents for a
two-dimensional, random Lorentz gas. In Sec.\ III, we calculate the
Lyapunov exponents through $O(\tilde{n}^{2})$ and obtain the constants
$A_1$ and $B_1$, defined above. We conclude the paper with some
discussions in Sec.\ IV. We note here that the structure of the
calculation presented in this paper is based on a thorough and
intricate mathematical formalism, which is summarized in the Appendix.

\section{Lyapunov exponents of the random Lorentz gas in two
dimensions: General Theory}

\noindent The random Lorentz gas consists of point particles of mass
$m$, moving among a random array of fixed scatterers. In two
dimensions, the scatterers are hard disks of radius $a$. We will
consider two versions of the model: the Lorentz gas with overlapping
scatterers, in which each configuration of scatterers a priori has
equal probability (hence no weighting according to the amount of free
volume left to the point particles!), and the Lorentz gas with
non-overlapping scatterers, in which scatterers cannot overlap each
other, but each configuration of scatterers satisfying this constraint
a priori is equally likely. For a system with $N$ scatterers in a
two-dimensional volume $V$ the number density of scatterers is
$n=N/V$, and at low density $\tilde{n}  \ll 1$. There is no
interaction between point particles. These particles, therefore,
travel freely during flights between collisions with the
scatterers. The collisions between a point particle and a scatterer
are instantaneous, specular and elastic. During flights, the equations
of motion of a point particle are
\begin{eqnarray}
\dot{{\bm r}}\,=\,{{\bm v}}\,=\,\frac{{\bm p}}{m}, {\hspace{1cm}}
\dot{{\bm p}}\,=\,m\dot{{\bm v}}\,=\,0\,.
\label{e2.1}
\end{eqnarray}
At a collision with a scatterer, the post-collisional position and
velocity, ${\bm r}_{+}$ and ${\bm v}_{+}$ of the point particle are
related to its pre-collisional position and velocity, ${\bm r}_{-}$
and ${\bm v}_{-}$, by
\begin{eqnarray}
{\bm r}_{+}\,=\,{\bm r}_{-}\,, {\hspace{1cm}} {\bm v}_{+}\,=\,{\bm
v}_{-}\,-\,2\,({\bm v}_{-}\cdot\hat{\bm \sigma})\,\hat{\bm
\sigma}\,. \label{e2.2}
\end{eqnarray}
Here $\hat{\bm \sigma}$ is the unit vector in the direction from the
center of the scatterer to the point of collision (see Fig.\
1). Notice that the dynamics generated by Eqs.\
(\ref{e2.1}-\ref{e2.2}) keep the speed of the point particle constant
at $v$. The instantaneous velocity direction of the particle is given
by $\hat{\bm v}(t)={\bm v}(t)/v$.

\begin{figure}[!h]
\centerline{\includegraphics[width=1.5in]{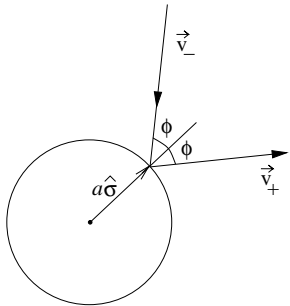}}
\caption{Collision between a point particle and a scatterer.}
\end{figure}

\begin{figure}[h]
\begin{center}
\includegraphics[width=2in]{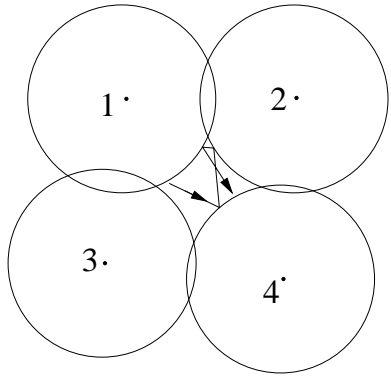}
\vspace{3mm}
\caption{A situation for the Lorentz gas with overlapping scatterers,
wherein the point particle cannot leave the enclosure formed by the
four scatterers. }
\label{figenclosure}
\end{center}
\end{figure}

The Lorentz gas with circular scatterers is a hyperbolic dynamical
system. Due to the convex nature of the collisions with the
scatterers, typically, the distance between two infinitesimally close
trajectories in phase space, $[{\bm r}(t),{\bm v}(t)]$ and $[{\bm
r}+\delta{\bm r}(t),{\bm v}(t)+\delta{\bm v}(t)]$, increases
exponentially with time. There are two non-zero Lyapunov exponents,
which sum to zero. We denote the positive Lyapunov exponent by
$\lambda_{+}$.  Without any loss of generality, one can characterize
the time evolution of the separation between two nearby trajectories
by that of its projection onto ${\bm v}$-space. The positive Lyapunov
exponent then can be defined as
\begin{eqnarray}
\lambda_{+}\,=\,\lim_{T\rightarrow\infty}\,\lim_{|\delta{\bm
v}(t_0)|\rightarrow0}\,\frac{1}{T}\,\ln\,\frac{|\delta{{\bm
v}}(t_{0}\,+\,T)|}{|\delta{{\bm v}}(t_{0})|}\,.
\label{e2.4}
\end{eqnarray}
The non-overlapping random Lorentz gas is generally supposed to be
ergodic, even in the infinite-system limit (although we do not know of
any proof of this).  Therefore this definition of $\lambda_{+}$ should
be (almost) independent of the choice of the initial point of the
trajectory. For the overlapping Lorentz gas the situation is slightly
more subtle: in this model, if the volume becomes large enough,
ergodicity will be broken. There will always be finite enclosures
formed by three or more scatterers, from which a point particle cannot
escape if it is trapped inside initially (see Fig.\
\ref{figenclosure}). However, as long as the density of scatterers is
below a critical percolation density, in the infinite system limit
there will always be a single unbounded percolating region on which
the motion of the point particles is diffusive on large time and
length scales. We will be interested in the Lyapunov exponents of such
particles only (even though particles moving inside an enclosure will
also exhibit a positive Lyapunov exponent).

As an alternative to Eq.\ (\ref{e2.4}), one can also choose to measure
the separation of two nearby trajectories in ${\bm r}$-space --- after
all, $\delta{\bm v}(t)=\displaystyle{\frac{d\delta{\bm
r}(t)}{dt}}$. In this representation the positive Lyapunov exponent
becomes
\begin{eqnarray}
\lambda_{+}\,=\,\lim_{T\rightarrow\infty}\,\lim_{|\delta{\bm
r}(t_{0})|\rightarrow0}\,\frac{1}{T}\,\ln\,\frac{| \delta{{\bm
r}}(t_{0}\,+\,T)|}{|\delta{{\bm r}}(t_{0})|}\,.
\label{e2.3}
\end{eqnarray}

When calculating $\lambda_{+}$ using Eq.\ (\ref{e2.4}), one may take
advantage of the feature that the separation in velocity space between
the two trajectories undergoes a change {\it only\,} at the collisions
with the scatterers. If the point particle suffers $k$ collisions
between time $t_{0}$ and $t_{0}+T$, then \cite{vBLD_pre_98}
\begin{eqnarray}
\lambda_{+}\,=\,\lim_{T\rightarrow\infty}\,\lim_{|\delta{\bm
v}(t_{0})|\rightarrow0}\,\frac{k}{T}\,\frac{1}{k}\,\sum_{i=1}^{k}\ln\frac{|\delta{\bm
v}_{i+}|}{|\delta{\bm v}_{i-}|}\,,
\label{e2.9}
\end{eqnarray}
where, $\delta{\bm v}_{i-}$ and $\delta{\bm v}_{i+}$ are respectively
the pre- and the post-collisional separation in velocity space between
the two trajectories at the $i$-th collision.

On the other hand, to obtain the time evolution of $|\delta{{\bm
r}}(t)|$, one may introduce another dynamical quantity, called the
radius of curvature, and defined as
$\rho=v\displaystyle{\frac{|\delta{\bm r}(t)-\hat{\bm
v}(t)\cdot\delta{\bm r}(t)|} {|\delta{\bm v}(t)-\hat{\bm
v}(t)\cdot\delta{\bm v}(t)|}}$
\cite{vBD_prl_95,vBLD_pre_98,D_cup_book}. It characterizes the
divergence between neighboring trajectories and in two dimensions
it may be defined as the distance of the actual positions on a
pair of such trajectories to their mutual intersection point (with
$\rho$ positive if this point is found in the past).  In terms of
the radius of curvature the positive Lyapunov exponent may be
expressed as \cite{Sinai_rms_70},
\begin{eqnarray}
\lambda_+\,=\,\lim_{T\rightarrow\infty}\,\frac{v}{T}\int_{t_{0}}^{t_{0}\,+\,T}\,\frac{d\,t}{\rho(t)}.
\label{e2.5}
\end{eqnarray}

During a free flight, the equation of motion for $\rho$ is given by
\cite{D_cup_book}
\begin{eqnarray}
\dot{\rho}\,=\,v\,.
\label{e2.6}
\end{eqnarray}
At a collision with a scatterer, the post-collisional radius of
curvature $\rho_{+}$ is related to the  pre-collisional radius of
curvature $\rho_{-}$ by \cite{vBD_prl_95,vBLD_pre_98,D_cup_book} :
\begin{eqnarray}
\frac{1}{\rho_{+}}\,=\,\frac{1}{\rho_{-}}\,+\,\frac{2}{a\cos\phi}\,,
\label{e2.7}
\end{eqnarray}
which is well-known from geometric optics. Here $\phi$ is the
collision angle, i.e, $\cos\phi\,=\,|\hat{\bm v}_{-}\cdot\hat{\bm
\sigma}|\,=\,|\hat{\bm v}_{+}\cdot\hat{\bm \sigma}|$ (see Fig.\
1). Equations (\ref{e2.6}) and (\ref{e2.7}), for a given initial
condition $[{\bm r}(t_0),{\bm v}(t_0)]$ and a fixed spatial
arrangement of the scatterers, can be solved together to obtain
$\rho(t)$ as a function of ${\bm r}(t_0)$ and ${\bm v}(t_0)$. Consider
a set of trajectory pairs generated from the same reference trajectory
by starting tangent trajectories with $\delta{\bm r}(t_0)=0$ and
$\delta{\bm v}(t_0)=\delta{\bm v}_0$ at a range of initial times
$t_0$. One easily convinces oneself that the radius of curvature along
the trajectory rapidly approaches a limiting value, $\rho({\bm r},{\bm
v})=\lim_{t_0\rightarrow-\infty}\rho({\bm r},{\bm v},t_0)$, with
$\rho({\bm r},{\bm v},t_0)$ the radius of curvature at the phase point
$({\bm r},{\bm v})$ for a trajectory bundle starting out from $[{\bm
r}(t_0),{\bm v}(t_0)]$. Combining Eqs.~(\ref{e2.6}) and (\ref{e2.7})
one finds that as a function of decreasing $t_0$ $\rho({\bm r},{\bm
v},t_0)$ can be expressed as a rapidly converging continued fraction.
In Eq.~(\ref{e2.5}) $\rho(t)$ therefore can be replaced by $\rho({\bm
r}(t),{\bm v}(t))$, independent of initial conditions, and, assuming
ergodicity, one may replace the long time average in Eq.~(\ref{e2.5})
by an equilibrium average \cite{vBD_prl_95,vBLD_pre_98,D_cup_book}.
Besides an average over initial position and velocity of the point
particle this involves an average over all allowed configurations of
the scatterers. In the sequel we will denote it as
\cite{vBD_prl_95,vBLD_pre_98,D_cup_book}
\begin{eqnarray}
\lambda_{+}\,=\,\left\langle\frac{v}{\rho}\right\rangle\,.
\label{e2.8}
\end{eqnarray}

A very useful way of rewriting the positive Lyapunov exponent uses
the relationship \be \frac{|\delta{\bm v}_+|}{|\delta{\bm
v}_-|}=\frac{\rho_-}{\rho_+}, \label{deltavdeltarho} \ee with the
$\rho$'s defined again as the values on a trajectory starting in
the infinite past. Combining this with Eq.~(\ref{e2.5}) one finds
that for an ergodic system the positive Lyapunov exponent may be
expressed as \be
\lambda_+=\nu_{\text{c}}\left\langle\ln\frac{\rho_-}{\rho_+}\right\rangle_{\text{coll}},
\label{lambda+rhorho} \ee where $\nu_{\text{c}}$ is the average
collision frequency, which is the inverse of the mean free time
between collisions $\tau_{\text{c}}$, and the average now runs
over the equilibrium distribution of collision configurations.
Using Eq.~(\ref{e2.7}) one may rewrite this further as \be
\lambda_+=\nu_{\text{c}}\left\langle\ln
\frac{a\cos\phi+2\rho_-}{a\cos\phi}\right\rangle_{\text{coll}}.
\label{lambda+rhophi} \ee

\section{The Density Expansion of the Positive Lyapunov Exponent}

We now present an analysis  for calculating the positive Lyapunov
exponent, $\lambda_{+}$, up to $O(\tilde{n}^{2})$, inclusive. As
mentioned before, this is done in two steps. In the first step, we
assume that the collisions of the point particle with the scatterers
are uncorrelated, implying that the probability density for a
collision with collision angle $\phi$ (see fig.\ 1) at all times is
given as \be p_{\text{coll}}(\phi)=\nu_{\text{c}} \cos\phi.  \ee We
calculate the resulting contribution to the positive Lyapunov exponent
up to $O(\tilde{n}^2)$, by means of both the velocity deviation method
and the radius of curvature method and show that the two agree.

In the second step, we calculate the corrections $\delta \lambda_+$
resulting from the non-overlap condition (in the overlapping scatterer
model) and from correlated collisions, i.e.\ either collisions of the
point particle with a scatterer it has collided with before, since the
initial time $t=t_0$, or collisions with a collision rate that is
slightly enhanced by the available information on the absence of other
scatterers on the preceding free path.

\subsection{Uncorrelated collision approximation}

\begin{figure}[!h]
\begin{center}
\includegraphics[width=3in]{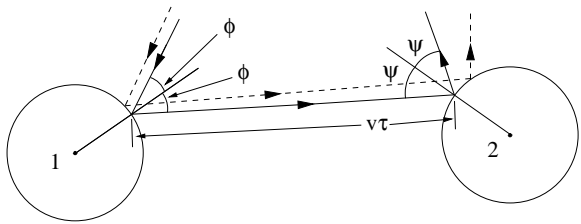}
\vspace{3mm}
\caption{A picture of a collision sequence. The solid and the dashed
lines are two trajectories of the point particle, infinitesimally
displaced from each other. Scatterer 1 and scatterer 2 are denoted by
1 and 2 respectively.}
\label{fig2}
\end{center}
\end{figure}

In the velocity deviation method our starting point is Eq.\
\ref{lambda+rhophi}. First of all we note that the contribution
$\nu_{\text{c}}\langle\ln(1/\cos\phi)\rangle_{\text{coll}}$ is
given {\em exactly} for all densities by \bea
\nu_{\text{c}}\langle\ln(1/\cos\phi)\rangle_{\text{coll}}&=&-\nu_{\text{c}}\int_{-\pi/2}^{\pi/2}
d\,\phi \frac{\cos\phi}2 \ln\cos\phi\nonumber\\
&=&\nu_{\text{c}}(\ln 2 -1). \label{lambdacos} \eea So, it remains
to calculate the contribution
$\displaystyle{\nu_{\text{c}}\left\langle\ln \frac
{a\cos\phi+2\rho_-}{a}\right\rangle_{\text{coll}}}$. To
investigate this, consider fig.\ \ref{fig2}. The collision angles
with the scatterers 1 and 2 are denoted as $\phi$ and $\psi$
respectively. From Eq.\ \ref{e2.6} it follows that the radii of
curvature of 2 and 1 just before respectively just after the
collision of the light particle are related as \be
\rho_{-}^{({\text 2})}=\rho_{+}^{({\text 1})}+v\tau, \label{rho-}
\ee and from Eq.\ \ref{e2.7} we find that \be \rho_{+}^{({\text
1})}=\frac{a\cos\phi}2-
\frac{\displaystyle{\left(\frac{a\cos\phi}2\right)^2}}
{\displaystyle{\rho_{-}^{({\text1})}+\frac{a\cos\phi}{2}}}.
\label{rho+} \ee
>From the last equation we see that at low density of the
scatterers $\rho_{+}^{(i)}$ is of order $a$ and typically by an
order of $\n$ smaller than the free path $v\tau$.  Therefore, in
our calculations of Lyapunov exponents the contributions from
$\rho_{+}^{({\text 1})}$ in Eq.\  (\ref{rho-}) only will
contribute to higher density corrections and not to the leading
terms.  For similar reasons the last term in Eq.\  (\ref{rho+})
may be ignored completely in our calculations, as it will not
contribute to corrections of order smaller than $\n^2$. The term
still to be calculated may then be written as
$\nu_{\text{c}}\displaystyle{\left\langle \ln
  \frac{2v\tau+a(\cos\phi+\cos\psi)}a \right\rangle_{\text{coll}}}$.  This has to be
averaged over the distribution of $\phi$, $\psi$ and $\tau$, which are
all three assumed to be independent. The distribution of $\tau$, under
the assumption of constant collision frequency, becomes a simple
exponential. Therefore the resulting expression for the Lyapunov
exponent becomes. \bea \lambda^{(\text{U})}_{+} &=&\nu_{\text{c}}
\int_{-\pi/2}^{\pi/2}d\,\phi\frac{\cos\phi}2
\int_{-\pi/2}^{\pi/2}d\,\psi \frac{\cos\psi}2
\int_0^{\infty}d\,\tau\,\nu_{\text{c}} e^{-\nu_{\text{c}}\tau} \ln
\frac{2v\tau+a(\cos\phi+\cos\psi)}a\nonumber\\ &=&\nu_{\text{c}}
\left[\ln \frac{v}{a\nu_{\text{c}}}+1-{\cal
    C}\right]
    -\frac{\pi\nu_{\mbox{\scriptsize c}}}{2}\tilde{n} \left[\ln\frac{a\nu_{\mbox{\scriptsize c}}}{4v}+\frac16+{\cal C}\right]
\label{lambdaU} \eea where ${\cal C}=0.5772\cdots$ equals Euler's
constant. To rewrite this in terms of the scatterer density, one has
to express the collision frequency in terms of the latter. For
non-overlapping scatterers the overall collision frequency depends on
the configuration of scatterers, but on average it equals \be
\nu_{\text{c}}^ {( \text{ov}) }=2nav.
\label{nuov} \ee for all densities.  However, this is only true if
one averages over all initial points for the point particle, that is,
points inside the enclosures as well as on the percolating region.  At
low densities the collision frequency of a particle inside an
enclosure will be much higher than the overall average value, so the
collision frequency of a particle in the percolating area, which we
are really interested in, has to be smaller.  Fortunately, these
corrections are at most of relative order $\n^2$ compared to the
leading term (because it takes at least three scatterers to make an
enclosure), so in our present analysis they may be ignored. In the
case of non-overlapping scatterers the collision frequency, for all
densities, follows immediately from the ratio between the sum of the
circumference of all the scatterers to the available free volume for
the point particle, as \be
\nu_{\text{c}}^{(\text{nov})}=\frac{2nav}{1-n\pi a^2}.
\label{nunov}
\ee Substituting these expressions into Eq.\ (\ref{lambdaU}), we find
that $\lambda^{(\text{U})}_{+} $ for the respective cases behaves as
\bea \lambda^{(\text{U})(\text{ov})}_{+} &=&2nav \left[\ln \frac
1{2\n}+1-{\cal C}+\frac{\pi} 2 \n\left(\ln\frac 2 \n-\frac 1 6-{\cal
C}\right)\right],
\label{lambdaUov}\\ \lambda^{(\text{U})(\text{nov})}_{+} &=&2nav\left[\ln \frac 1{2\n}+1-{\cal C}+\frac{\pi} 2
\n\left(3\ln\frac 1 \n -\ln 2 +\frac{11}6- 3{\cal C}\right)\right].
\label{lambdaUnov}
\eea

In the radius of curvature method, $\lambda^{(\text{U})}_{+}$ is
calculated by taking into account precisely the same dynamical
aspects and approximations that have been used in the velocity
deviation method above. The identity of the results may be
established immediately by rewriting the integral in Eq.\
(\ref{e2.5}) as a sum of integrals between subsequent collisions
of the point particle.  By using Eqs.\ (\ref{e2.6}) and
(\ref{rho-}) one immediately finds that the positive Lyapunov
exponent may be expressed as \be
\lambda_+=\nu_{\text{c}}\left\langle \ln \frac{\rho_-^{(i+1)}}
{\rho_+^{(i)}}\right\rangle_{\text{coll}}, \ee which is obviously
equivalent to Eq.\ (\ref{lambda+rhorho}).

The approximation that the collisions suffered by the point particle
are uncorrelated is taken into account by the use of an extended
Lorentz-Boltzmann equation (ELBE) --- defined below --- for the
distribution function $\tilde{f}({\bm r},{\bm v},\rho,t)$ of the
moving particle in $({\bm r},{\bm v},\rho)$-space at low density of
scatterers. This function describes the probability density of finding
the moving particle at position ${\bm r}$ with velocity ${\bm v}$ and
radius of curvature $\rho$, at time $t$, averaged over all  allowed
configurations of scatterers (remember that for given configuration of
scatterers $\rho$ is a uniquely defined function of ${\bm r}$ and
${\bm v}$). Since the speed of the particle is constant we may replace
${\bm v}$ by the angular variable $\theta$ describing the angle
between ${\bm v}$ and the $x$-axis. In equilibrium the distribution
function $f({\bm r},\theta,\rho)$ is a function of the radius of
curvature only and the extended Lorentz-Boltzmann equation takes the
form \cite{vBD_prl_95,vBLD_pre_98,D_cup_book}
\begin{widetext}
\begin{eqnarray}
v\,\frac{\partial f(\rho)}{\partial \rho}&=&\nu_{\text{c}}
\int_{-\frac{\pi}{2}}^{\frac{\pi}{2}}\,d\phi\frac{\cos\phi}2\,\Theta\left(\frac{a\cos\phi}{2}\,-\,\rho\right)
\int_{0}^{\infty}d\rho'\,\,\delta\left(\rho\,-\,\frac{\rho'a\cos\phi}{a\cos\phi\,+\,2\rho'}\right)
\,\,f(\rho')\,-\,\nu_{\text{c}} f(\rho)\,,
\label{e3.6}
\end{eqnarray}
\end{widetext}
\noindent
with $\Theta(x)$ the unit step function. In this case, the
distribution function ${f}({\bm r},\theta,\rho,t)$ satisfies the
normalization condition
\begin{eqnarray}
\displaystyle{\int_{0}^{\infty}d\rho\,{f}({\bm r},\theta,\rho,t)}
\equiv {F}({\bm r},\theta,t)\,=\,\frac{1}{2\pi V}\,.
\label{e3.7}
\end{eqnarray}
Here ${F}({\bm r},\theta,t)$ is the distribution function of the point
particle in $({\bm r},\theta)$ space.

The quantity $\rho'$ in the argument of the $\delta$-function in Eq.\
(\ref{e3.6}), is the pre-collisional radius of curvature that produces
a post-collisional radius of curvature $\rho$.  From Eq.~(\ref{rho-})
and the fact that the free path $v \tau$ for most inter-collision paths
is of the order $\tilde{n}^{-1}$, it follows that in most cases $\rho'
\gg a$.  As a result of this one may to leading order in the density
simplify Eq.~(\ref{e3.6}) by using the approximation
\cite{vBD_prl_95,vBLD_pre_98,D_cup_book}.
\begin{eqnarray}
\delta\left(\rho\,-\,\frac{\rho'a\cos\phi}{a\cos\phi\,+\,2\rho'}\right)\,
\approx\,\delta\left(\rho\,-\,\frac{a\cos\phi}{2}\right)\,.
\label{e3.8}
\end{eqnarray}
The solution of Eq.\ (\ref{e3.6}) obtained with this simplification,
will be denoted as $f^{(0)}(\rho)$. The positive Lyapunov exponent to
lowest order in the density of scatterers, $\lambda^{(0)}_{+}$, is
obtained by using the distribution function $f^{(0)}(\rho)$ in Eq.\
(\ref{e2.8}) to calculate the ensemble average. We express the
solution of Eq.\ (\ref{e3.6}), $f(\rho)$ as a power series expansion
in $\tilde{n}$ as
\begin{eqnarray}
f(\rho)\,=\,f^{(0)}(\rho)\,+\,\delta\!f^{(0)}(\rho)\,+\,.\,.\,.
\label{e3.9}
\end{eqnarray}
and obtain the expression of $\lambda^{(\text{U})}_{+}$ as
\begin{eqnarray}
\lambda^{(\text{U})}_{+}\,=\,2\pi V
v\int_{0}^{\infty}d\rho\,\,\frac{f^{(0)}(\rho)\,+\,\delta\!f^{(0)}(\rho)}{\rho}\,.
\label{e3.10}
\end{eqnarray}
We specifically point out here that to calculate
$\delta\!f^{(0)}(\rho)$, one cannot use the approximation in Eq.\
(\ref{e3.8}) any longer. The function $\delta\!f^{(0)}(\rho)$ may be
calculated \cite{Panja_unpub} by applying a successive approximation
scheme to Eq.\ (\ref{e3.6}). Density corrections to $\lambda_0$,
contained in $\lambda_+^{\text{(U)}}$, result from both
$f^{(0)}(\rho)$ and $\delta f^{(0)}(\rho)$. This procedure for
calculating these corrections is algebraically more cumbersome than
the velocity deviation method. Nevertheless, it equally leads to the
results of Eqs.\ (\ref{lambdaUov}-\ref{lambdaUnov}). However, for
systems that are not, on average, spatially homogeneous (e.g.\ a
Lorentz gas with open boundaries) the velocity deviation method
becomes more complicated and in fact the ELBE method is to be
preferred \cite{latzetal}.

In the second step of our calculations we consider corrections due to
correlations between collisions. From Eqs.\ (\ref{lambda+rhophi}),
(\ref{rho-}) and (\ref{rho+}) it becomes clear that there are two
types of corrections that have to be accounted for:
\begin{itemize}
\item The approximation of $\rho_-$ by $v\tau+a\cos\phi/2$.
\item Deviations from the exponential
$\nu_{\text{c}}\exp(-\nu_{\text{c}}\tau)$ of the distribution of the
free flight time $\tau$ between collisions.
\end{itemize}
Through order $\n^2$ the first type of corrections are only
important in case $v\tau$ is comparable to $a$, that is if two
scatterers with which the light particle collides are close to
each other. If this is the case, the light particle may recollide
an arbitrary number of times with either scatterer and at each of
the collisions the actual value of $\rho_-$ has to be used instead
of the uncorrelated collision approximation. This will be
discussed further in the next subsection.  The second type of
corrections to order $\n^2$ only occur for the case of
non-overlapping scatterers.  They will be investigated in the
next-to-next subsection.

\subsection{Corrections due to nearby scatterers\label{sec3b}}
\begin{figure}[h]
\includegraphics[width=0.3\linewidth]{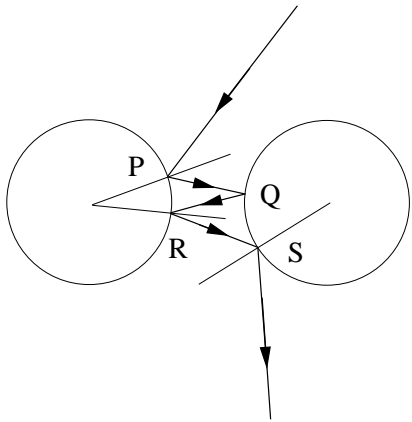}
\caption{An example scattering sequence involving multiple collisions
with two scatterers.}
\label{mult}
\end{figure}

In case two scatterers have a mutual distance comparable to their
radius $a$, a first collision of the point particle with either of
them with fairly large probability will lead to recollision sequences
such as the one shown in fig.\ \ref{mult}.

For the first two collisions, at $P$ and $Q$, the approximation of
$\rho_-$ by $v\tau+a\cos\phi/2$, with $\tau$ and $\phi$ the
preceding free time respectively the collision angle at the
previous collision, is still adequate. For all subsequent
collisions $\rho_-$ has to be calculated from Eqs.\ (\ref{e2.7})
and (\ref{rho-}). This gives rise to a correction \be
\delta\lambda_+^{(\text{ns})}=\nu_{\text{c}}\displaystyle{\left\langle\ln
\frac{2\rho_-
+a\cos\psi}{2v\tau+a(\cos\psi+\cos\phi)}\right\rangle_{\text{coll}}}
\label{deltalambdanb} \ee to the positive Lyapunov exponent. The
average has to be taken over all two-scatterer configurations with
nearby scatterers and over all points on the circumference of
scatterer 1 where a collision may occur (so in the overlapping case
points covered by 2 are excluded) and over all allowed values of the
collison angle $\psi$, weighted again by $\cos \psi/2$. In fact only
those parameters that correspond to a recollison with 1 give rise to
non-vanishing  contributions at the relevant orders of density. For
the overlapping scatterer model the initial configurations of the
scatterers do include overlapping ones; for the non-overlapping model
such configurations are excluded. We evaluated the averages
numerically, with the results \bea
\delta\lambda_+^{(\text{ns})(\text{ov})}&=&-0.38\,(2nav)\n\,\quad{\mbox{and}}\nonumber\\
\delta\lambda_+^{(\text{ns})(\text{nov})}&=&-0.28\,(2nav)\n\,.
\label{deltalambdans}
\eea Notice that the probabilities that these collision sequences are
interrupted by other scatterers can clearly be ignored at the orders
of the scatterer density we are considering presently.

\subsection{Corrections to the free-path length distribution\label{sec3c}}
The assumption of a constant collision frequency, independent of the
past, is not entirely correct.  For the case of overlapping scatterers
this only becomes manifest when one considers corrections of relative
order $\n^2$. So for our present purposes we may ignore this. For
non-overlapping scatterers corrections already show up at the order
$\n$.  These appear in two forms: shortly after a collision the
probability for a next collision is reduced by a {\em shadowing
effect}: many scatterer locations that would give rise to such a
collision are forbidden by the non-overlap condition. This is
compensated by an {\em anti-shadowing effect} at long times; the
absence of scatterers in a strip of width $2a$ around the free path of
the point particle increases the probability of finding a scatterer
with which this particle will collide.  Obviously, the combined
influence of these two effects on the overall collision frequency has
to vanish, but there is a shift of collision probability towards
longer times, shifting the free-path length distribution equally to
somewhat larger values of the free path.

\begin{figure}
\centerline{\includegraphics[width=2.1in]{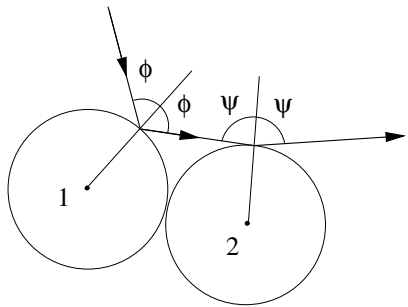}}
\caption{For given scattering angles $\phi$ and $\psi$ the free path
length between subsequent collisions has a minimal value
$b(\phi,\psi)$. This minimum is determined by the condition that the
two scatterers touch each other.}
\label{fig3}
\end{figure}

\begin{figure}
\centerline{\includegraphics[width=2.1in]{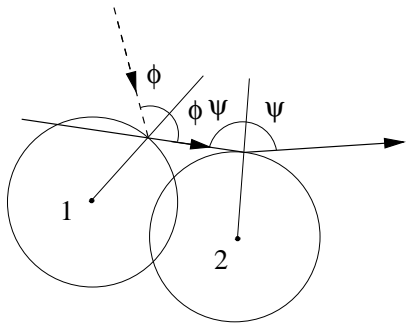}}
\caption{After a long free flight the probability that a collision
will occur like the one shown with scatterer 2, is slightly enhanced
by the fact that no scatterers like 1 can be in the way. If the
collision angles are $\psi$ and $\phi$ respectively, the distance over
which the absence of 1 influences the probability of finding 2 is just
$b(\phi,\psi)$.}
\label{figantishadow}
\end{figure}

The shadowing effect is illustrated in fig.\ \ref{fig3}. For
subsequent scattering angles $\phi$ and $\psi$ the minimal path length
between the collisions is given by \be
b(\phi,\psi)=a\left[\sqrt{4-(\sin\phi-\sin\psi)^2}-\cos\phi-\cos\psi\right].
\label{b}
\ee
\noindent
The contributions from shorter path lengths have been counted
erroneously in  $\lambda_+^{(\text{U}) (\text{nov})}$ and should be
subtracted.  This gives rise to a correction of the form \bea
\delta\lambda_+^{(\text{shad})(\text{nov})}&=&-(2nav)^2
\,\,\angintphi\,\,\,\,\,\angintpsi\int_0^{b(\phi,\psi)/v} d\tau
\ln\left(\frac {2v\tau} a+\cos\phi+\cos\psi\right),\nonumber\\
&=&-0.0976\,(2nav)\,\n\,.
\label{deltalambdashad}
\eea
\noindent
The anti-shadowing effect occurring for long free times, is
illustrated in fig.\ \ref{figantishadow}.
For a collision with collision angle $\psi$ there is with certainty no
obstruction from a scatterer that would have given rise to a previous
collision with scattering angle $\phi$ within a preceding distance
$b(\phi,\psi)$.  This enhances the collision frequency by an amount of
\bea \delta\nu_{\text{c}}^{(\text{lt})}=\nu_{\text{c}}\,\,
\angintphi\,\,\,\angintpsi\,\,\, n a\, b(\phi,\psi)\,&=&\,
\frac{\pi+18\sqrt{3}-32}{24}\,\,\nu_{\text{c}}\,,\\
&=&\,0.0966\cdots\nu_{\text{c}}\nonumber
\label{deltanu}
\eea One should also take into account, however, that the probability
for a collision at long times is enhanced somewhat by the initial
suppression of the collision frequency through the shadowing effect.
The combined effects lead to a long term collision probability density
of the form \be
p_{\text{coll}}(t)=(\nu_{\text{c}}+\delta\nu_{\text{c}}^{(\text{lt})})
e^{-(\nu_{\text{c}}+\delta\nu_{\text{c}}^{(\text{lt})})(t-t_d)}.
\label{pcoll}
\ee Here $t_d$ is the average delay time due to the shadowing
effect. It is given by \bea t_d&=&\frac 1
v\,\,\angintphi\,\,\angintpsi b(\phi,\psi)
\label{td}
\eea and related to $\delta\nu_{\text{c}}^{(\text{lt})}$ by \be
\delta\nu_{\text{c}}^{(\text{lt})}=\nu_{\text{c}}^2 t_d.  \ee The
ensuing correction to the positive Lyapunov exponent now may be
obtained as\cite{poslyap} \bea
\delta\lambda_+^{(\text{as})(\text{nov})}&=&\nu_{\text{c}}\int_0^{\infty}
dt\left\{(\nu_{\text{c}}+\delta\nu_{\text{c}}^{(\text{lt})})
e^{-(\nu_{\text{c}}+\delta\nu_{\text{c}}^{(\text{lt})})(t-t_d)}-
\nu_{\text{c}} e^{-\nu_{\text{c}} t}\right\}\ln \frac {2vt}
a,\nonumber\\ &=&\delta\nu_{\text{c}}^{(\text{lt})}\,\n\left[\ln \frac
1 {2\n}-{\cal C}\right].
\label{deltalambdaas}
\eea Collecting the contributions to $\delta\lambda_+$ from Eqs.\
(\ref{lambdaUov}), (\ref{lambdaUnov}), (\ref{deltalambdans}),
(\ref{deltalambdashad}) and (\ref{deltalambdaas}), we obtain our final
result, \bea \delta\lambda_+^{(\text{ov})}&=&2nav \left[\ln \frac
1{2\n}+1-{\cal C}+\frac{\pi} 2 \n\left(\ln\frac 2 \n-\frac 1 6-{\cal
C}-0.242\right)\right]\quad\mbox{and}\\
\delta\lambda_+^{(\text{nov})}&=&2nav\left[\ln \frac 1{2\n}+1-{\cal
C}+\frac{\pi} 2 \n\left\{3\ln\frac 1 \n -\ln 2 +\frac{11}6- 3{\cal
C}-0.240+\frac{\pi+18\sqrt{3}-32}{12\pi}\left(\ln \frac 1 {2\n}-{\cal
C}\right)\right\}\right].\label{final}  \eea

\section{Discussion}

In this paper, we have calculated the Lyapunov exponents of a
two-dimensional, random Lorentz gas at low densities up to
$O(\tilde{n}^2)$ in the density of scatterers. This calculation was
carried out in two parts: in the first part we assumed that subsequent
collisions between the light particle and one of the scatterers are
uncorrelated. In the second part we calculated the effects of
correlations between collisions and, in the case of non-overlapping
scatterers, those resulting from the non-overlap condition. The
effects of repeated recollisions of the light particle with two
scatterers that are close to each other, were calculated numerically,
as well as the contribution from
$\delta\lambda_+^{(\text{shad})(\text{nov})}$. All other contributions
to the positive Lyapunov exponent were obtained analytically. For the
sake of brevity and understanding, we have presented the method in
this paper in fairly intuitive terms, as opposed to the more formal
mathematical structure upon which all calculations have been based
originally \cite{Kruis_thesis}.  A short summary of this formalism, is
included in the Appendix at the end of this paper.

We argued that our analysis covers all density correction to the
Lyapunov exponent through order $\tilde{n}^2$. We have no rigorous
proof of this, but it is easy to support this claim by simple power
counting arguments.

The formalism was developed in the context of calculating the Lyapunov
exponents of a two-dimensional Lorentz gas at low densities up to
$O(\tilde{n}^2)$, but the method itself is not limited to the
calculation of the Lyapunov exponents alone -- it can be used for a
density expansion of other dynamical quantities of a two-dimensional
Lorentz gas as well. We note that the formalism as well as the
somewhat intuitive analysis in this paper, in principle can be
extended so as to obtain expressions for the Lyapunov exponents up to
higher orders in the density of scatterers. Though it would be very
desirable to have reliable results for all densities, we expect a
systematic extension of the present methods will be very laborious and
complicated. Perhaps it will be possible to find non-systematic, but
still accurate approximations, such as Enskog equations or ring
kinetic equations in the kinetic theory of transport processes. In
that case the present theory may be very valuable in providing
guidelines and testing criteria at moderate densities.

A more modest goal would be an extension of the present analysis to
systems out of equilibrium.  Here one may think first of all of
systems with escape and systems under the action of a driving field
combined with a gaussian thermostat. This should certainly be doable.

\noindent ACKNOWLEDGEMENTS: it is a great pleasure dedicating this
paper to Carlo Cercignani. We do not aim for the rigor of his papers
but hope he will enjoy the systematics of the analysis presented.

We thank Bob Dorfman for many helpful discussions and Michiel
Meeuwissen for his help with the numerical integrations of Eq.
(\ref{deltalambdans}).

HvB acknowledges support by the Mathematical Physics program of
FOM and NWO/GBE and DP acknowledges support by the Physics
Department of the University of Maryland during the preparation of
his doctoral thesis, on which this paper has been based in part.

\appendix\section*{Appendix}

Here we want to illustrate how the results obtained in this paper may
be obtained in a systematic way from a diagrammatic analysis of the
dynamics of the Lorentz gas.

Starting point is the binary collision expansion (BCE), explained
for the Lorentz gas, respectively the hard sphere gas in
\cite{vanLeeuwen,EDHV}. The dynamics of the point particle is
expressed as a series of convolution products involving sequences
of alternating free streaming and binary collision operators. The
free streaming operators describe the ballistic motion of the
point particle between two collisions with a scatterer. The binary
collision operators consist of a real and a virtual part. Both of
them check for the conditions of a collision to be satisfied. The
real collision operator implements the velocity change of the
point particle, as given by Eq.\ (\ref{e2.2}). The virtual
collision operator leaves the velocity unchanged but multiplies by
$-1$. It always produces corrections to simpler terms in the
binary collision expansion, which ignore the collision and let the
point particle move freely through the scatterer.

The binary collision expansion is built by starting from a free
streaming operator ignoring all collisions occurring in reality. The
first corrections consist of terms containing one collision operator,
with the real one generating a path that contains one collision with a
specific scatterer and the virtual one subtracting the free streaming
terms for all initial configurations from which free streaming indeed
gives rise to a collision with that given scatterer. Next,
two-collision terms give rise to paths with two real collisions and
correction terms with either one real and one virtual, or two virtual
collision terms, and so on.

The non-overlap condition between point particle and scatterers may be
accounted for at the initial time, but, as shown by Van Beijeren and
Ernst \cite{vBE} it has great advantages to postpone this, as much as
possible, until the times of the first collision of the point particle
with any specific scatterer. This may be done efficiently by choosing,
in each term of the BCE, the {\em first} collision operator with any
given scatterer, unless it is the last collision with this scatterer
as well, to be a $\overline{T}$ operator, rather than a $T$ operator,
used for all remaining collisions. The difference between these two
binary collision operators is that the $\overline{T}$ operator counts
virtual collisions at the instant that the point particle leaves the
scatterer, whereas the $T$ operator counts them at entrance.

\begin{figure}[h]
\centerline{\includegraphics[width=4in]{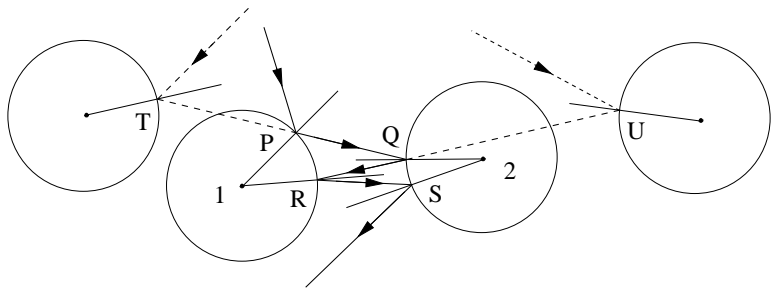}}
\caption{Cancellations between real and virtual collision diagrams
in binary collision expansion for repeated collisions between the
point particle and two scatterers separated by distances of
$O(a)$.} \label{multplus}
\end{figure}
The uncorrelated collision approximation is reproduced by keeping in
the BCE exactly those terms in which each scatterer appears at most
once. A good way of rearranging these terms is separating each of the
binary collision operators into its real and virtual part, and taking
together all events that have the same sequence of real collision
operators. Terms with arbitrary number of virtual collisions between
two subsequent given real collisions occurring at times $t_j$ and
$t_{j+1}$, sum to yield the damping term
$\exp[-\nu_{\text{B}}(t_{j+1}-t_j)]$, with $\nu_B=2nav$ the low
density limit of the collision frequency. For the case of
non-overlapping scatterers one may extend the analysis so as to
replace the low density collision frequency $\nu_{\text{B}}$ by the
full collision frequency $\nu_{\text{c}}$. To this end one has to
replace the collision operators in the BCE by sums of operators
representing the product of the collision operator with the pair
correlation function of the point particle and a scatterer, at
contact. The value of the latter just equals $(1-\pi\tilde{n})^{-1}$,
thus reproducing Eq.\ (\ref{nunov}). Notice that also the particles
responsible for the static correlations at collisions occur in these
BCE terms at just one single point in time.

Next, consider correlations due to the same scatterer being
present at more than one collision. As in the main text, we
distinguish between nearby cases, where all scatterers involved
are within mutual distances of the order $a$, and events in which
at least one of these distances is of the order of the mean free
path, $\ell=a/(2\n)$. In the nearby case each additional scatterer
reduces the weight by an additional factor proportional to $\n$,
so in the approximation considered here, we may restrict ourselves
indeed to events involving just two scatterers. As stated in the
main text, both for the overlapping and for the non-overlapping
case we obtain contributions from events in which there is at
least one recollision with either of the two scatterers. As an
example, consider an event of the same type as exposed in Fig.\
\ref{mult}, but now with the collisions preceding virtual
collisions at $\text{P}$ or $\text{Q}$ added (see fig.\
\ref{multplus}). Consider first the contributions to Eq.\
(\ref{deltavdeltarho}) from collisions at $\text{R}$. For all
BCE-events containing a collision at $\text{S}$ one has exact
cancellation from the contribution with a real and that with a
virtual collision at $\text{S}$. So we only need consider events
for which the collision at $\text{R}$ is the last one. The event
with just one preceding collision, at $\text{Q}$, was included
already at the uncorrelated collision approximation. The event
where a virtual collision at $\text{P}$ is followed by real ones
at $\text{Q}$ and $\text{R}$ subtracts the contribution from the
event with subsequent real collisions at $\text{T}$, $\text{Q}$
and $\text{R}$ in the uncorrelated collision approximation. The
contribution with subsequent real collisions at $\text{P}$,
$\text{Q}$ and $\text{R}$ finally, adds the actual contribution
from the collision at $\text{R}$. Next, consider the contributions
from the collision at $\text{S}$. The event with a virtual
collision at $\text{Q}$ subtracts the uncorrelated event with
collisions at $\text{U}$, $\text{R}$ and $\text{S}$. The events
with a virtual collision, respectively no collision at $\text{P}$,
followed by real collisions at $\text{Q}$, $\text{R}$ and
$\text{S}$, cancel each other and finally again, the event with
all collisions real gives the actual contribution. These
reasonings are easily generalized to collision sequences of
arbitrary length. They immediately lead to the result of Eq.\
(\ref{deltalambdanb}).

\begin{figure}[h]
\centerline{\includegraphics[width=0.8\linewidth]{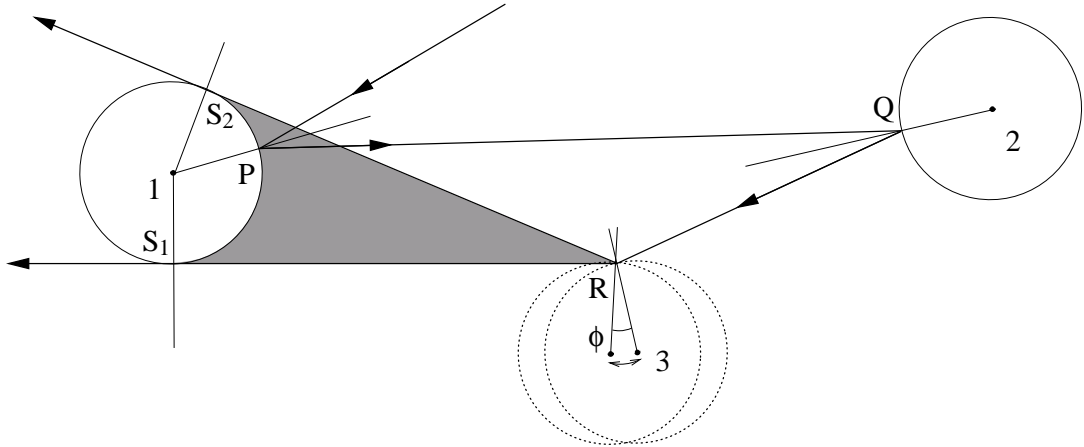}}
\caption{Limitation of the collision angle to $O(a/\ell)$ for
recollision events involving multiple scatterers that are
separated by distances of $O(\ell)$. Notice that the trajectories
between scatterers 1 and 3 are confined within the shaded grey
area, which limits the variation of the collision angle at R
within a range $\Delta\phi\sim a/\ell$.} \label{ring}
\end{figure}
In the non-overlapping case the contributions from consecutive
collisions with mutually overlapping scatterers need special
consideration (in the overlapping case such configurations play no
special role). In the diagrammatic representation of the BCE the
overlap gives rise to a non-vanishing Mayer-bond between the
overlapping scatterers and collision events like the one shown in
Fig.\ \ref{fig3} correspond to sections of diagrams with two different
$T$-operators connected by a Mayer-bond between the two scatterers
involved. Either of the two collisions may be real or virtual. The
case of two real operators leads to the correction
$\delta\lambda_+^{\text{(shad)(nov)}}$ of subsection \ref{sec3b}. The
case of a real collision followed by a virtual one is responsible for
the delay time $t_d$ defined in Eq.\ (\ref{td}). A virtual collision
followed by a real one is responsible for the enhancement
$\delta\nu_{\text{c}}^{\text{(lt)}}$ of the collision frequency as
specified in Eq.\ (\ref{deltanu}). The diagram segment with two
virtual collisions accounts for the enhancement by
$\delta\nu_{\text{c}}^{\text{(lt)}}$ of the damping factor in the
survival probability during free flight.  Finally, events involving
recollisions between overlapping scatterers cancel exactly against the
same events without the Mayer-bond between the two scatterers
included. In subsection \ref{sec3b} this was accounted for by
restricting the contributions from recollision events involving two
nearby scatterers to non-overlapping configurations in the
non-overlapping model.

\begin{figure}[ht]
\centerline{\includegraphics[width=0.7\linewidth]{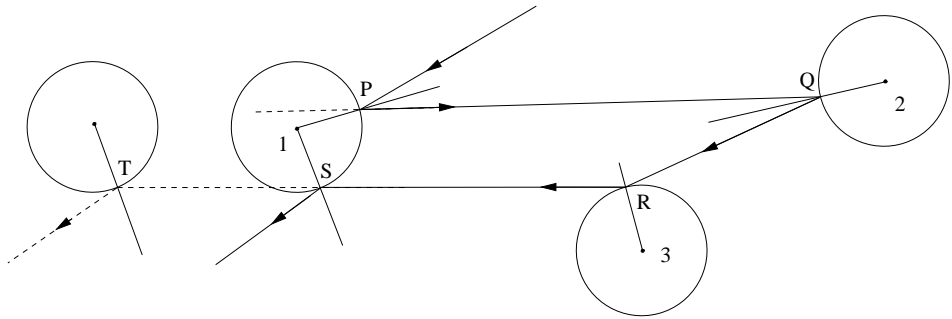}}
\caption{Cancellations between real and virtual collision diagrams in
binary collision expansion for repeated collisions between the point
particle and multiple scatterers separated by distances of $O(\ell)$.}
\label{ringb}
\end{figure}
Recollisions where the point particle returns to a scatterer from
a distance of the order of the mean free path, restrict the
allowed collision angle at the preceding collision to an angular
range of the order $a/\ell\sim \n$ (see Fig.\ \ref{ring}).
Therefore, at the level of corrections to $\lambda_+$ restricted
to order $\n^2$, no further restrictions may be imposed on the
free flight lengths and collision angles of the intermediate
collisions and all of these lengths typically are of order $\ell$.
Now consider, as an example, contributions to $\delta\lambda_+$
resulting from the recollision at $\text{S}$ in Fig.\ \ref{ringb}.
According to Eq.\ (\ref{lambda+rhorho}) these come from the
averages of $\ln\displaystyle{\frac{\rho_-}{\rho_+}}$ at
$\text{S}$, in case the recollision is real, or at $\text{T}$, in
case it is virtual. In either case, the contribution in which the
collision at $\text{P}$ is virtual, cancels that where the latter
is real, through the order of $\n$ considered. The same holds true
for the contributions from all later collisions. Collisions before
the recollision of course are not influenced by it at all. Hence,
we conclude that corrections to $\lambda_+$ from recollision
events involving distances of the order $\ell$ are absent through
order $\n^2$ and we were justified in ignoring these in subsection
\ref{sec3c}.


\begin{thebibliography}{00}
\bibitem{commentSinai} To be precise: this holds under the conditions
that there is no "infinite horizon", i.e.\ there are no trajectories
of infinte length that never hit a scatterer, and, in the case of
overlapping scatterers, that the space between the scatterers is
percolating (extends to infinity).
\bibitem{LB} Since we use the full collision frequency in our
equations this should perhaps rather be called a Lorentz-Enskog
equation. Substituting $\nu_{\text{c}}=2nav$ one indeed recovers the
Lorentz Boltzmann equation of \cite{vBD_prl_95}.
\end{thebibliography}
\end{document}